# Patterning Sn-based EUV resists with low-energy electrons

*Ivan Bespalov[†,\*], Yu Zhang[†], Jarich Haitjema[†], Rudolf M. Tromp[‡,§], Sense Jan van der Molen[‡], Albert M. Brouwer[†,||], Johannes Jobst[‡,\*], and Sonia Castellanos[†,\*]*

[†]Advanced Research Center for Nanolithography, Science Park 104, 1098XG Amsterdam, The Netherlands

[‡]Kamerlingh Onnes Laboratory, Leiden University, Niels Bohrweg 2, 2333 CA Leiden, The Netherlands

[§]IBM T. J.Watson Research Center, 1101 Kitchawan Road, P.O. Box 218, Yorktown Heights, NY 10598, USA

[||]University of Amsterdam, van't Hoff Institute for Molecular Sciences, P.O. Box 94157, 1090 GD Amsterdam, the Netherlands

ABSTRACT. Extreme Ultraviolet (EUV) lithography is the newest technology that will be used in the semiconductor industry for printing circuitry in the sub-20 nm scale. Low-energy electrons (LEEs) produced upon illumination of resist materials with EUV photons (92 eV) play a central role in the formation of the nanopatterns. However, up to now the details of this process are not well understood. In this work, a novel experimental approach that combines Low-Energy Electron Microscopy (LEEM), Electron Energy Loss Spectroscopy (EELS), and Atomic Force Microscopy (AFM) is used to study changes induced by electrons in the 0-40 eV range in thin films of molecular organometallic EUV resists known as tin-oxo cages. LEEM-EELS spectroscopic experiments were used to detect surface charging upon electron exposure and to estimate the electron landing energy. AFM post-exposure analyses revealed that irradiation of the resist with LEEs leads to the densification of the resist layer associated to carbon loss. The same chemical processes that yield densification render the solubility change responsible for the pattern formation in the lithographic application. Remarkably, electrons as low as 1.2 eV are able to induce chemical reactions in the Sn-based resist. Based on the thickness profiles resulting from LEE exposures in the 3-48 mC/cm$^2$ dose range, a simplified reaction model is proposed where the resist undergoes sequential chemical reactions, yielding first a sparsely cross-linked network, followed by the formation of a denser cross-linked network. This model allows us to estimate a maximum reaction volume on the initial material of 0.15 nm$^3$ per incident electron in the energy range studied, which means that less than 10 LEEs per molecule on average are needed to turn the material insoluble and thus render a pattern. The results presented in this work give novel and highly relevant insights into the chemical efficiency of LEEs of different energies in state-of-the-art EUV resist materials.

**INTRODUCTION**

Traditional lithography based on Deep Ultraviolet (DUV) light has reached its resolution limit and now requires complex processing steps to print ever smaller and denser components in integrated circuits.[1] Yet, as the miniaturization of electronic components in computer chips continues in order to keep up with Moore's law, novel nanopatterning technologies are necessary to attain a cost-effective high volume manufacturing. Among all nanopatterning approaches, Extreme Ultraviolet Lithography (EUVL) is the most promising candidate to reach the targeted sub-20 nm resolution, by employing a much shorter wavelength (13.5 nm) than it is used in current DUV lithography (193 nm).[2] One of the biggest challenges in the establishment of EUVL as the new workhorse of the semiconductor industry lies in the interaction of the high-energy (92 eV), extreme ultraviolet (EUV) radiation with the photoresist material. Conventional polymer-based photoresists designed for DUV lithography offer relatively low EUV photon absorption, which limits their performance.[3] Therefore, the search for new materials that can absorb an optimal amount of EUV light and render high-quality nanopatterns is essential for EUVL technology.[4–6]

Among the variety of materials that are being investigated for EUVL applications, metal-organic materials - also called inorganic resists - are considered the most promising. Their main advantage is that the incorporation of metallic elements enhances EUV absorptivity.[7] In particular, Sn-containing materials have attracted much attention as they can yield nanopatterns at relatively low doses.[8–10] Yet, a lack of detailed understanding of the chemical processes occurring upon the absorption of EUV photons hinders the rational design of efficient resists. When an EUV photon is absorbed by the resist, primary and secondary electrons with energies in the 0-80 eV range are produced.[11,12] These electrons play a central role in the chemical transformations that photoresists undergo. Specifically, they can induce molecular bond scissions,[13,14] which change the photoresist structure and thus its solubility properties, thereby enabling pattern formation.[11,15–21] However, very few studies of the electron energy-dependence of these processes have been performed up to date.[15,17,22–24]

Gaining knowledge on which electrons induce more significant changes in EUV photoresists is of high relevance both from a fundamental and an applied point of view. Mainly, the efficiency of electron-induced reactions contributes to the overall sensitivity of the photoresists.[18,19] At the same time, the so-called electron blur in the final nanopattern –the maximum distance away from the photon absorption point where electrons induce solubility changes– depends on the electron mean free path.[25] The latter is the average distance that an electron travels between scattering events and it has an inelastic and an elastic component. Accurate experimental values for the mean free paths of electrons below 100 eV are scarce and only recently it has been experimentally shown that they strongly depend on electron energy and on the material.[26] Therefore, for EUV lithography, understanding interactions of low-energy electrons with photoresist materials and the energy-dependence of those interactions presents an essential contribution to estimate, and eventually control, the efficiency of the photoresist as well as the lateral blur of patterns produced by low energy electrons, which in the last instance determines the resolution of the printed features.

In the present work we use Low-Energy Electron Microscopy (LEEM) to expose thin films of a Sn-based EUV resist with low-energy electrons within the 0-40 eV energy range, which is representative for the secondary electrons generated upon EUVL. We use Electron Energy Loss Spectroscopy (EELS) to determine with accuracy the energies of the electrons that imping the photoresist, correcting for surface charging effects that result from the poorly conducting character of the material. Next, using Atomic Force Microscopy (AFM), we study the electron-induced structural changes as a function of electron energy and exposure dose and relate them to the changes in the solubility properties of the material. These experiments allow us to estimate the average reaction volume per incident electron as a function of electron energy. Similarly, we estimate a "chemical efficiency" of LEEs in the 0-40 eV range in terms of number of electrons needed per molecule to render the material insoluble.

**RESULTS AND DISCUSSION**

We study films of tin-oxo cages, a molecular material referred to as TinOH, where the OH stands for the two hydroxyl counter ions[27,28] (**Fig. 1a**). This compound is a Sn-based material that has proven to be a promising EUV resist.[8,10,29,30] The mechanism responsible for the solubility change of TinOH promoted by EUV photons was proposed in previous works.[8–10,31] Here, we investigate how low-energy electrons directly induce changes in the solubility properties of this material as a function of electron energy and dose within a relevant energy window for EUVL, i.e. 0-40 eV.[32–34,15] The design of our LEEM experimental setup allows us to evaluate the effect of LEEs on the photoresist using *in-situ* and *ex-situ* approaches. In the *in-situ* approach, the interaction of LEEs with the photoresist is monitored using LEEM-based EELS (**Fig. 1b**). The *ex-situ* approach consists of exposure to low-energy electrons, followed by AFM analysis both before and after a development step is applied to the resist layer (**Fig. 1c-d**).

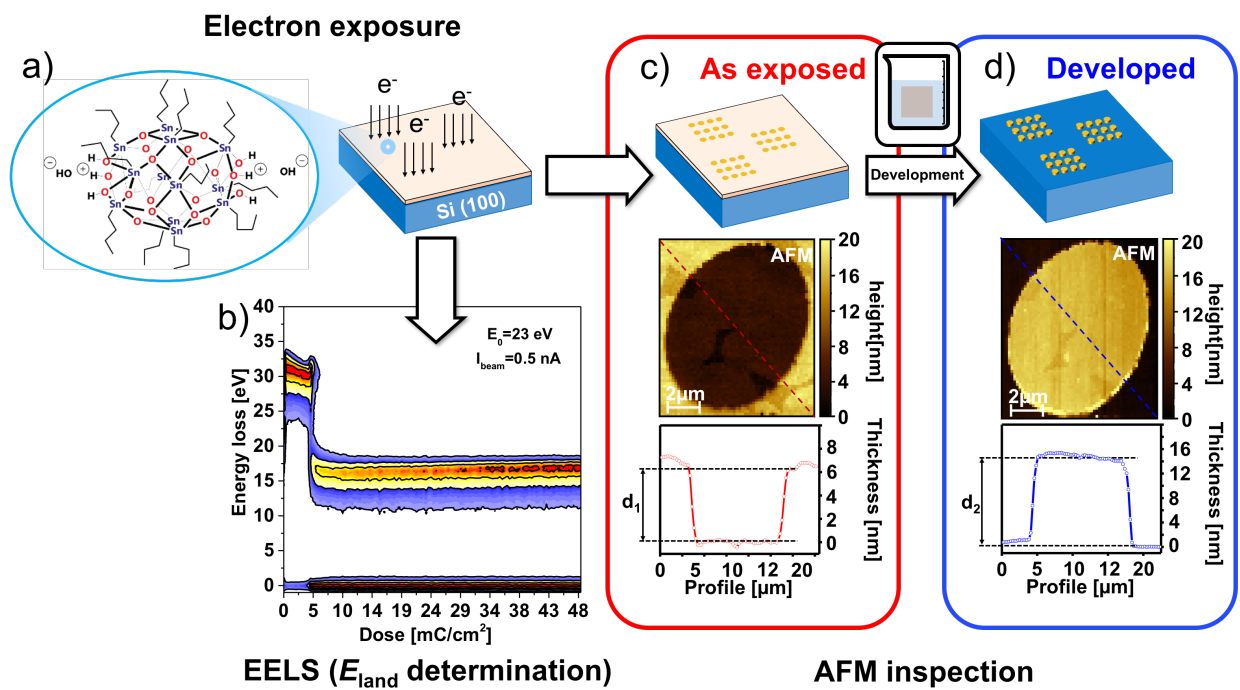

**Figure 1.** (a) Chemical structure of tin-oxo cage with hydroxide counter ions (TinOH). (b) Example of a transient electron energy loss (EELS) spectrum during electron exposure. (c), (d) Inspection by AFM of electron-induced changes in a thin film (20 nm) of TinOH exposed to electrons of $E_{land}$ = 15.8 eV and exposure dose of 12 mC/cm$^2$ in LEEM. (c) The exposed area is clearly visible in the AFM image on the "as exposed" film before development. The difference in thickness between exposed and unexposed areas is shown below in the profile line scan along the red dashed line. (d) AFM image of the same area shown in (c) after development. The thickness of the insoluble material left after development is shown below as a height profile line scan along the blue dashed line.

### *In-situ* EELS experiments: surface charging

When a poorly conducting resist layer is exposed to low-energy electrons, the resulting surface charging can severely affect the electron/resist interaction energy. This phenomenon has been studied in thin films of PMMA,[35] and is crucial for an accurate understanding of the electron-exposure experiments. To quantify the dynamic charging effects in the present experiment, EELS spectra were recorded during electron exposure for different primary electron beam energies ($E_0$) in the 0-40 eV range. $E_0$ is defined as the potential difference between the electron gun (-15 keV) and the potential applied to the sample ($V_s$), corrected by the work function difference $\Delta\Phi$ between electron emitter and sample ($E_0$ = -15 keV + $eV_s$ + $\Delta\Phi$). In reference 36, we have shown that the width of the EELS spectrum, i.e, the difference between the zero-loss peak and the secondary electron cut-off, provides a direct measurement of the electron landing energy ($E_{land}$), i.e. the actual energy that the electrons have when they reach the surface of the sample. Given that TinOH is a poorly conducting material, $E_{land}$ is in general not equal to $E_0$, due to charging effects. Hence, we use the width of the EELS spectrum during exposure (**Fig. 1b**) in these experiments to determine $E_{land}$, i.e. the true interaction energy, independently.

To do so, for each particular setting of $E_0$, the time-evolution of the EELS spectrum was recorded during electron exposure, up to a dose of 56 mC/cm$^2$. Examples of such measurements are shown in **Fig. 1b** and **Fig. 2a-c**, which display the energy distribution of the electrons reflected and emitted by the sample as a function of exposure dose at a constant value of $E_0$. The y-axis thus corresponds to an energy "loss" scale, that is, the width of the EELS spectra that gives $E_{land}$, and the x-axis to dose, which is proportional to exposure time, i.e. dose = time × current density (**Fig. 1b**, **Fig. 2a-c**).

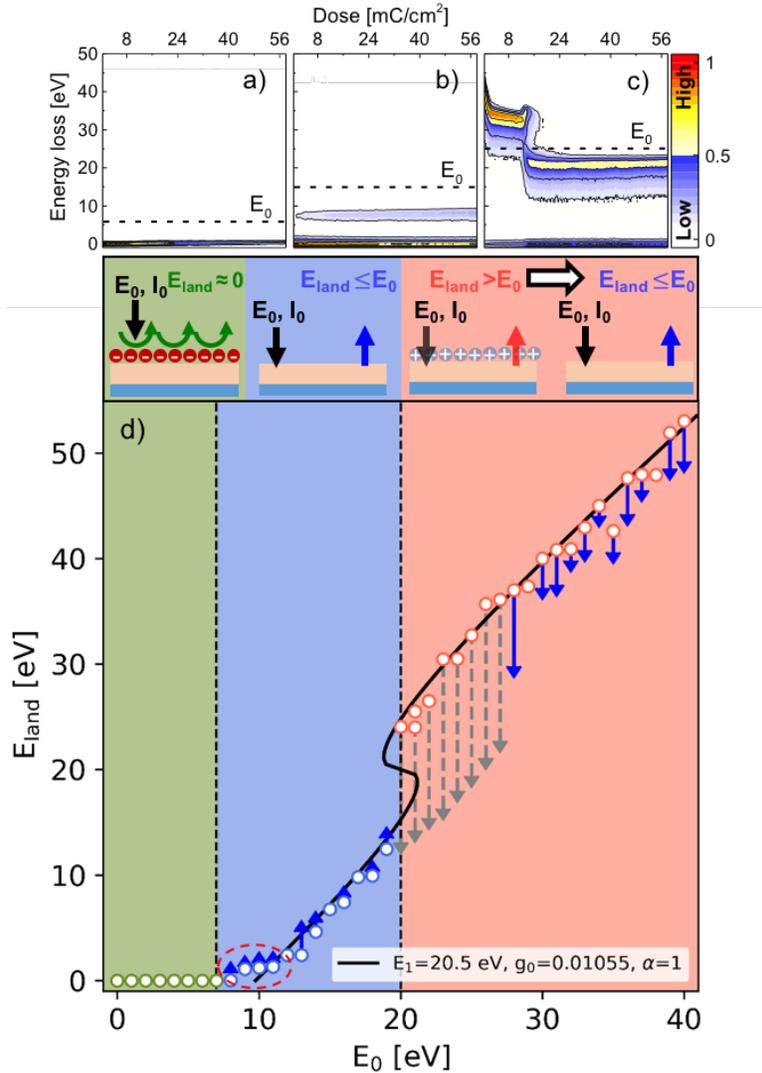

**Figure 2.** Effect of photoresist surface charging during electron exposure on $E_{land}$ measured by LEEM-EELS. **(a)-(c)** EELS spectra recorded during exposure of 20 nm thick TinOH films in LEEM for electron energies $E_0$ = 7 eV **(a)**, 16 eV **(b)**, and 26 eV **(c)**. **(d)** Dependence of the measured $E_{land}$ on $E_0$. White-filled circles represent the $E_{land}$ values directly after the start of the exposure (at dose = 0 in **a-c**). The solid line plots the fit of the experimental result using the cusp catastrophe equation developed in reference 35 for normalized conductance, $g_0$ = 0.01055, and the energy at which secondary electron emission equals the incident electron flux, $E_1$ = 20.5 eV. The arrows indicate the evolution of $E_{land}$ with increasing dose at each $E_0$ as a consequence of changes in the material over dose: blue arrows for $E_0$ values where $E_{land}$ shifts gradually, and dashed arrows for abrupt drops of $E_{land}$ as the secondary electron emission coefficient shifts from > 1 to < 1. The dashed red ellipse highlights the $E_0$ energies where $E_{land}$ fluctuates around zero

due to small fluctuations in the beam current. All measurements were performed at a constant beam current density $I_0$ = 0.017 nA/μm².

In **Fig. 2** it can be observed that three surface charging regimes can be distinguished in three different ranges of $E_0$. In the green area of **Fig. 2d** ($E_0$ = 0-7 eV, $E_{land} \approx 0$), the photoresist surface is charged negatively[37] and repels all incident electrons so that only the zero loss peak is observed in the EELS spectrum (**Fig. 2a**). In the blue area ($E_0$ = 7-19 eV, $0 < E_{land} < E_0 < E_1 = 20$ eV), the negative charge decreases due to increasing secondary electron emission, and incoming electrons now interact with the sample with energy $E_{land}$. In the red area ($E_{land} > E_0 > E_1 = 20$ eV), secondary electron emission coefficient is greater than unity and surface charging is positive. As a consequence, $E_{land}$ is higher than $E_0$. Chemical changes on the sample induced by the electrons lead to a decrease of the secondary emission coefficient (i.e. increase in $E_1$) over time (dose) as well as to an increase in the normalized conductance ($g_0$) of the film (see **Fig. S1** in Supporting Information). This induces a shift in $E_{land}$ during exposure, which is represented by arrows in **Fig. 2d**. The increase of $E_1$ with exposure leads to a shift of the blue/red boundary to higher values of $E_0$, and a decrease of the secondary electron emission coefficient to a value below 1 after a certain dose, which is accompanied by a sudden drop from $E_{land} > E_0 > E_1$ to $E_{land} < E_0 < E_1$ (**Fig. 2c**), represented with the dashed arrows in **Fig. 2d**. These observations are in agreement with previous results on PMMA, and can be quantitatively described by a so-called catastrophe theory.[35] The black S-shaped curve in **Fig. 2d** is a fit to the zero-exposure data based on this theory. More details about the theory and the evolution of the S-curve with electron exposure dose can be found in reference 35.

In the rest of this work we will use the measured landing energies, $E_{land}$, as shown in **Fig. 2d**, to define the energy of the incident electrons.

### *Ex-situ* AFM analysis: electron-induced densification and solubility changes

Post-exposure AFM inspection was used to monitor changes in the resist film thickness induced by electrons of different energies and doses, as well as for detecting changes in the solubility properties of the resist. For the latter purpose, the sample was immersed in a developer (2-heptanone/water mixture) that selectively dissolves the starting material but not the products formed upon exposure.[15] Examples of AFM images recorded before ("*as exposed*") and after development ("*developed*") are shown in **Fig. 1c** and **d**, respectively. In the "as exposed" sample (**Fig. 1c**), the dark ellipse reveals that the irradiated area undergoes a substantial thickness decrease (densification) with respect to the surrounding non-irradiated area. After development, the unexposed resist is washed away, leaving behind only the exposed areas that have turned insoluble due to electron-induced chemistry (**Fig. 1d**). To quantify the densification and the amount of insoluble material, the film thickness of the same exposed areas is measured before and after development by means of AFM and is plotted as a function of electron energy and dose in **Fig 3**.

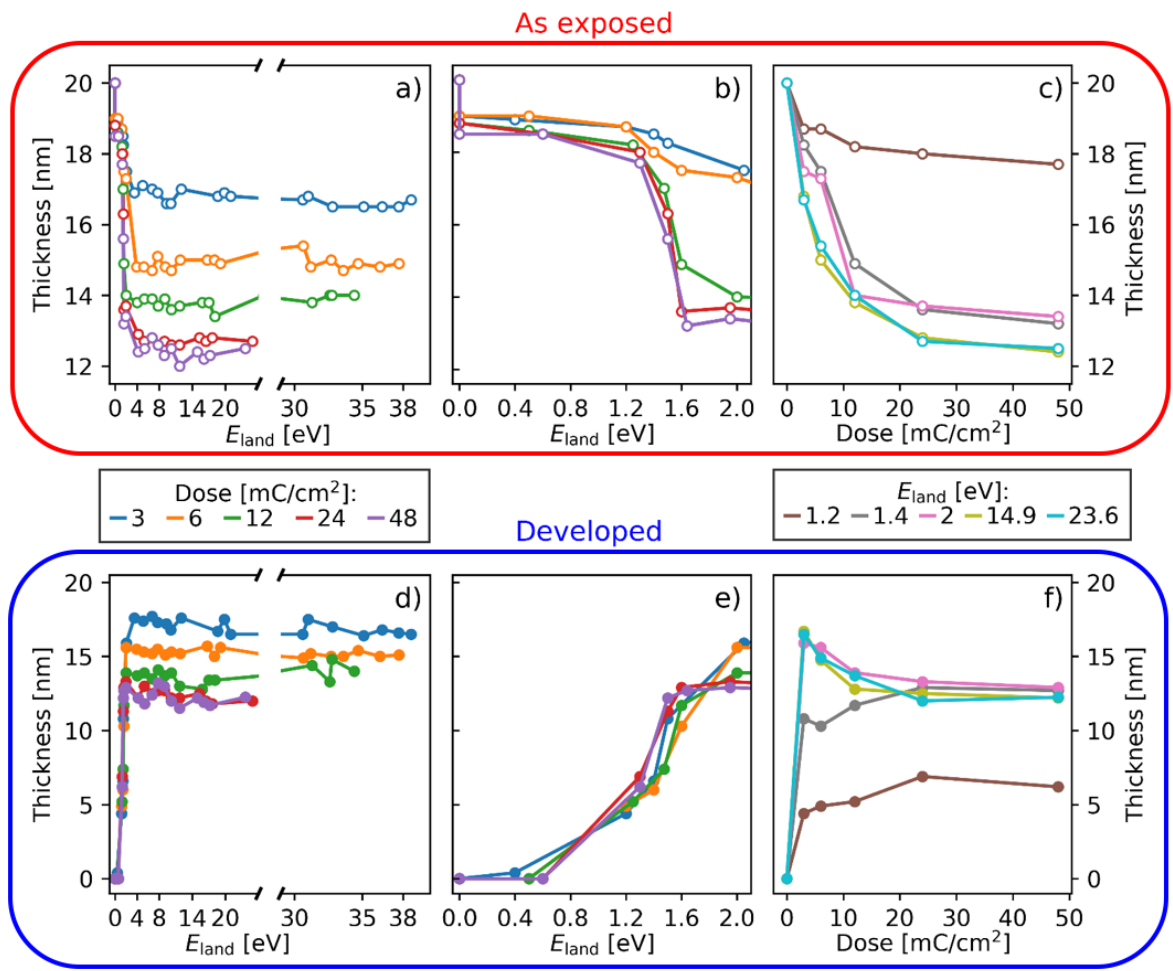

**Figure 3.** Thickness measured on 20 nm-thick TinOH resist layer as a function of electron energy ($E_{land}$) before development, **(a)** and zoomed-in plot **(b)**, and after development, **(d)** and zoomed-in plot **(e)**; and as a function of dose for some selected energies before **(c)** and after **(f)** development. Exposures performed using 0.017 nA/μm$^2$ incident electron beam current density.

The thickness of the exposed areas in the "as exposed" sample significantly decreases with increasing electron dose and energy (**Fig. 3a-b**). Given the low electron energy and current density, we dismiss beam-induced, direct evaporation of whole molecules as the origin of thickness loss and attribute it to a densification of the resist layer as a result of electron-induced chemical reactions. Such a densification has been also observed in layers of this material upon EUV exposure when they are over-exposed beyond the dose that renders the whole thin film insoluble.[29] A likely reaction mechanism for this process is discussed in detail below.

At each given dose, the film thickness decreases to the same value for all electrons with $E_{land} >$ 2 eV (**Fig. 3a**), which indicates that the densification induced by electrons in 4-36 eV range is rather similar. Yet, the densification increases with dose, reaching a maximum value at the highest dose used in our experiments, 48 mC/cm$^2$. The evolution of the densification is even clearer when thickness is plotted as a function of dose at a given $E_{land}$ (**Fig. 3c**). It appears that electrons of 0-1.4 eV in the measured dose range do not yield as much densification. Yet, it should be noticed that in the onset region where $E_{land}$ starts to deviate from zero (red-dashed circle in **Fig. 2d**), relatively small (10-20%) fluctuations in the incident electron current during electron exposure will have the effect of $E_{land}$ fluctuating around zero. Therefore, only a fraction of the incident electrons impinges on the sample and the actual dose "absorbed" by the material is lower than the incident one. Thus, the resist reactivity *appears* to be reduced, whereas in fact it is the dose that is reduced. Unfortunately, it is not possible at present to measure the exact dose reduction in this narrow energy window. For $E_{land} > \sim 2$ eV this effect no longer occurs, and exposure dose is unambiguous.

The thickness evolution due to electron exposure observed in the "as exposed" films (**Fig. 3a,b**) is mirrored in its "developed" version (**Fig. 3c,d**). During development, the unexposed material is washed away whereas the material in the irradiated areas remains, in line with the negative tone behavior previously reported for this resist.[10,29,30] This shows that the chemical changes that lead to thickness shrinking are also responsible for changes in the solubility of the material. As for the "as exposed" sample, for $E_{land} > 2$ the remaining thickness after development (**Fig. 3c**) does not vary significantly with the energy increase, for a given dose. However, the layer thickness decreases with increasing dose, reflecting the densification trends already observed in the undeveloped material.

In **Fig. 3d** the plots of the remaining "developed" thickness as a function of dose resemble the contrast curves commonly used in photolithography to evaluate resists sensitivity.[29] There, the minimum dose of light of a specific wavelength necessary to induce a solubility switch (from soluble to insoluble in the case of a negative tone resist) can be determined from the onset of the curve. In the present curves, the dose onset is below 3 mC/cm$^2$ in all cases. Importantly, even very low energy electrons (1.2 eV electrons) can already induce chemical reactions that yield changes in the solubility properties of the material.

Interestingly, for all curves resulting from exposure to electrons with $E_{land} \geq 2$ eV a maximum thickness value is reached at the lowest dose (3 mC/cm$^2$) before decreasing to an almost constant value for doses above 24 mC/cm$^2$. Such a profile indicates that at low doses an insoluble product with a relatively low degree of densification is formed which keeps reacting and densifying as the electron dose is increased. It is known that exposure of TinOH to DUV photons leads to Sn-C bond cleavage, which yields volatile products derived from butyl chains that outgas from the film.[8] Also, butyl fragments have been detected in electron-induced desorption experiments performed on similar Sn oxocages with 80 eV electrons.[38] Since the butyl chains represent up to a ~ 70% of the molecule volume,[27] (see **Fig. S2** in Supporting Information), the cleavage of butyl chains necessarily renders a significant decrease of film thickness. As reference, a complete transformation of the TinOH film (density in the crystalline form 1.84 g/cm$^3$, giving a molar volume of 1341.6 cm$^3$/mol) to pure SnO$_2$ (density in the crystalline form 6.95 g/cm$^3$, giving a molar volume of 38.8 cm$^3$/mol) would lead to a compaction of ~ 65%, i.e. from 20 nm to 7 nm.

Hence, we attribute the observed "as exposed" densification to carbon-loss reactions. In addition, for similar Sn-based materials, it has been proposed that the Sn-C bond cleavage yields active Sn-sites prone to form bonds with neighboring activated sites. This leads to the subsequent aggregation of the inorganic clusters and the creation of an insoluble network.[21] Given that TinOH has 12 carbon chains per molecule, we expect a gradual butyl cleavage and cross-linking of the inorganic residues with increasing EUV or LEE irradiation, resulting in an increasingly denser material. In addition, other reactions involving Sn-O bond cleavage might also occur to a certain extent, which would also have an impact on the densification of the material.

In order to relate the thickness curves in **Fig. 3** to chemical changes, we simplify this complex process of reactions in a model where two types of products (materials **B** and **C**) with different densities are formed in sequence from the original TinOH (material **A**), that is, through a two-step reaction **A→B→C**.

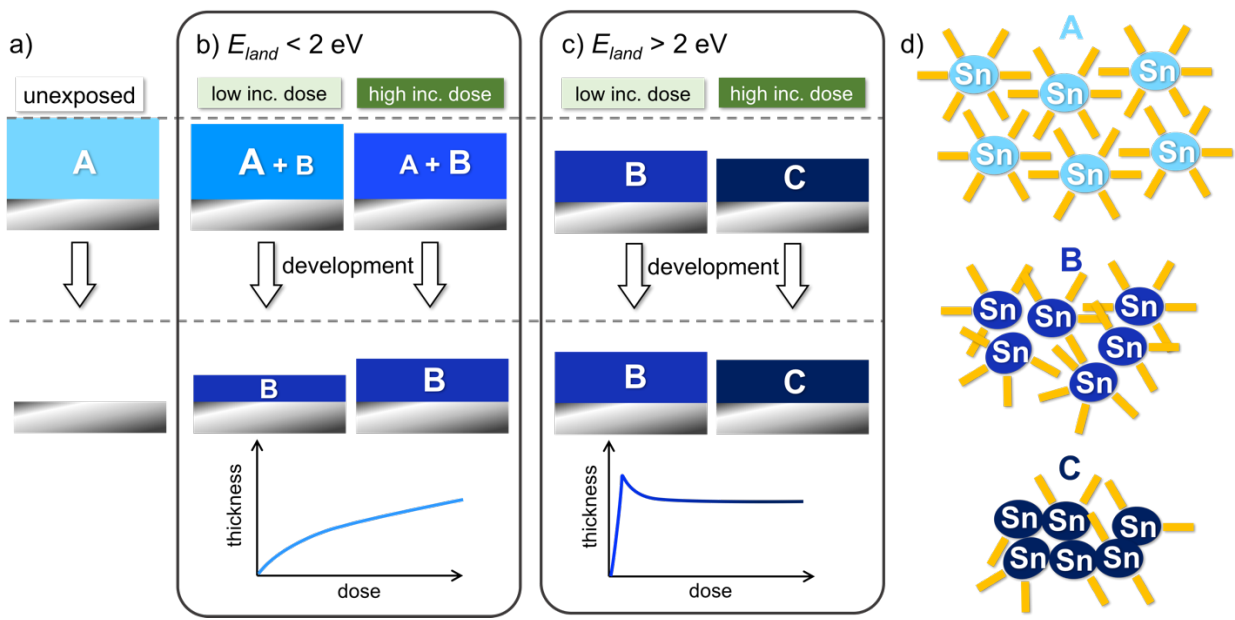

**Figure 4.** Scheme representing the densification of the TinOH material **A**. (a) Unexposed resist is removed completely during the development. (b) For electron exposure with $E_{land}$ < 2 eV only a small fraction of electrons impinging on the surface reach the material and only low conversion is attained. (c) For $E_{land}$ > 2 eV, as the incident (inc.) dose

increases, two consecutive reactions yield the insoluble products **B** (denser than **A**) and **C** (denser than **B**). **(d)** Schematic representation of the initial TinOH molecular material **A** and of the two insoluble networks **B** and **C**. Blue ellipses represent the Sn-based inorganic core and orange bars the butyl chains.

As schematically shown in **Fig. 4**, upon electron exposure the initial resist material **A** transforms first into an insoluble product **B** with higher density than **A**, due to few butyl-chain cleavage events, loss of the carbon chain by desorption, and cross-linking among the few molecular units through the "activated" Sn-sites (the ones that underwent Sn-C bond cleavage). Product **B** represents thus the mixture of relatively low weight oligomers that are cross-linked to a low degree. The **A→B** reaction requires low doses and the subsequent electrons can promote further carbon-loss and aggregation of the inorganic units leading to the **B→C** evolution, where **C** has an even denser structure due to further butyl-chain losses and potentially due to some structure collapse of the TinOH inorganic core (**Fig. 4b**). Product **C** in this model thus represents a network with a high degree of cross-linking among the Sn-containing core units and with a low carbon content. This transformation of the material with dose (carbon loss and enrichment of inorganic $SnO_x$ composition) results in a 9% increase of the material conductance ($g_0$) deduced from the fitting of the S-curves that result from surface charging evolution charging over dose (see **Fig. S2c** in Supporting Information)

For very low energy electrons (1.2 eV and 1.4 eV), the number of electrons that reach the resist at the given incident doses is not sufficient to transform all initial material **A** to the insoluble **B** or **C** and a mixture of mainly **A + B** is formed in the exposed areas. This is the same behavior observed in photoresists when the photon dose applied is not sufficient for a full conversion of the initial material into insoluble material.[39,40] In this under-exposed regime, the thickness of the exposed film is reduced in the development step since the remaining material **A** is dissolved (**Fig. 4a**).

The fraction of unreacted material **A** left in the exposed areas can thus be calculated by comparing the thickness of the film after compression (cf. 'as exposed' in **Fig. 3c**), which is a mixture of **A** and **B**, and the thickness remaining after development (cf. 'developed' in **Fig. 3f**), which consists only of the insoluble material (**B** at lower doses and **C** at higher doses). **Fig. 5a** shows the thickness lost during development as a function of exposure dose for some selected electron energies. It thus plots the conversion of the starting material **A** as a function of electron dose for electrons of different energies. It can be observed that a 3 mC/cm$^2$ dose of electrons with energies of 2 eV is already sufficient to transform the layer of initial material into the insoluble mixture that we identify as **B**. This dose, corresponding to an energy dose of 9 mJ/cm$^2$, is in the range of dose needed to transform the whole thickness of the material ($D_{100}$) when using EUV light: for a 40 nm film, ca. 50 mJ/cm$^2$ of EUV incident dose are required,[29] from which 38% is absorbed,[41] i.e. ca. 19 mJ/cm$^2$.

A likely mechanism for electron-induced chemistry in n-butyltin oxo cages is electron capture followed by the decomposition of the radical anion formed. This process, known as dissociative electron attachment, can be promoted by electrons of very low kinetic energies of the added electron, sometimes even 0 eV,[13,14,42–44] and are likely to occur in polarized bonds, such as metal-carbon bonds[42–44] like the Sn-C[38] in TinOH. Molecular quantum chemical calculations (see Supporting Information) support the notion that the radical anion that is formed after one electron gain is not stable. The Sn-C bond dissociation energy for this species is predicted to be only 0.4 eV by DFT calculation (B3LYP functional, Def2TZVP//LANL2DZ basis sets), much smaller than for the neutral molecule (predicted 2.3 eV, experimental for organotin compounds ~2.5 eV).[45]

While very low-energy electrons can decompose TinOH via electron attachment, electrons with higher kinetic energies can promote other mechanisms. Electrons that can transfer > 5 eV can bring the tin cage molecules to their electronically excited states, and, at energies > 7 eV, they can cause their ionization.[9] Both electronically excited and ionized tin cages undergo facile Sn-C bond cleavage.[8] This is because the lowest unoccupied molecular orbital in the neutral molecule (singly occupied in the radical anion and the lowest excited state) has Sn-C σ* anti-bonding character and, in the case of ionization, an electron is removed from the HOMO, which can be described as an Sn-C σ bonding orbital.[31] Thus, in all three cases, the Sn-C bond is significantly weakened.

Electrons with landing energies above the ionization energy are expected to generate secondary electrons in the bulk of the material. Yet, the exact secondary electron yield induced by incident electrons of different kinetic energies, as well as the energy distribution of those secondary electrons is not known for TinOH (and most other materials). The fact that the densification observed for the whole $E_{land}$ range of 4-36 eV is rather similar (**Fig. 3a**) while the secondary electron emission coefficient in EELS increases is intriguing. We speculate that this phenomenon might be related to similar mean free paths/penetration depths of the incident electrons in this energy range. The conversion of the full thickness into an insoluble material is evidenced by the fact that the exposed areas remain after development. Yet, the exposure to low energy electrons might lead to a gradient of the film chemical conversion from top to bottom (more chemical conversion and densification at the top) as a result of the short mean free paths of the incident electrons.[46] Unfortunately, the penetration depths/mean free paths of electrons in such low energy range in a complex material like TinOH are not known and cannot be determined with the present experiments. Therefore, the exact reason for the lack of energy dependence in the compression in the studied $E_{land}$ range remains uncertain.

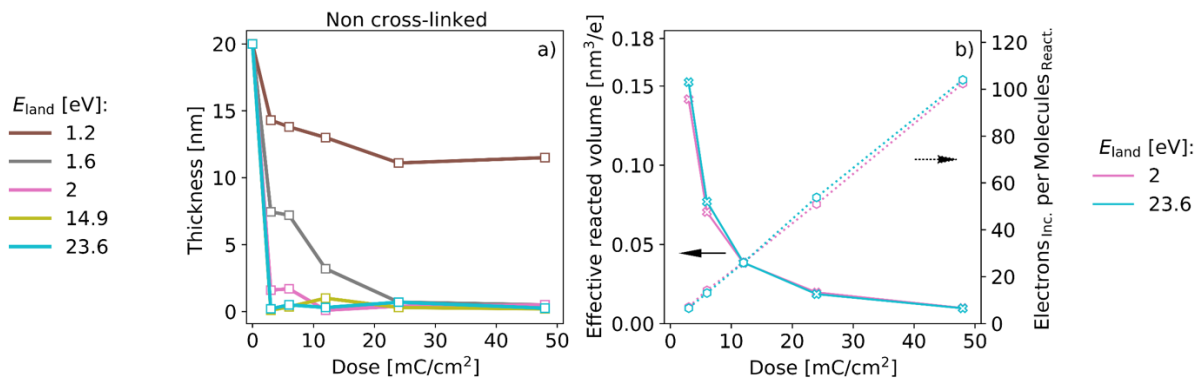

**Figure 5. (a)** Thickness of starting material **A** (non cross-linked material) dissolved in the development step for each dose of electrons of different energies ($E_{land}$). **(b)** Effective reacted volume per impinging electron (continuous line) and number of incident electrons per reacted molecule estimated for every exposure dose to electrons of selected energies.

From the exposed area (30 μm$^2$) and the difference between initial thickness of **A** (20 nm) and the thickness of unreacted **A** (as calculated in **Fig. 5a**) we calculated the volume of converted **A** over dose. Together with film density from the crystal structure (1.84 g/cm$^3$),[27] the TinOH molecular weight (2468.5 g/mol) and electron dose, we can then calculate how many electrons are needed per molecule on average to induce a solubility change as a function of electron energy (**Fig. 5b**). Initially less than 10 electrons per molecule are required regardless of the energy in the range between 2 and 34.2 eV. This number is in agreement with the number of secondary electrons involved in the solubility switch of the material when EUV photons are used, as estimated from previous works in the literature. From the EUV photon dose to render a 40 nm TinOH layer insoluble ($D_{100}$ = 50 mJ/cm$^2$)[29] mentioned above, it can be estimated that an average of 0.6 EUV photons are absorbed per molecule to yield the insoluble product **B**. And, although an experimental value for total electron yield per absorbed photon in TinOH has not been reported, for a very similar resist material photoelectron emission experiments indicated a yield of 2.3 secondary electrons per absorbed EUV photon[20] whereas a theoretical model found the best fitting to the experimental results in the assumption of 8 secondary electrons generated per absorbed photon.[21] The combination of these estimations from the literature, suggest that an average of 1.4 to 4.8 electrons per molecule could yield the insoluble network, which is within the range of our results.

Similarly, by dividing the reacted volume by the number of electrons an average reacted volume per incident electron can be estimated (**Fig. 5b**). The maximum obtained reaction volume per electron is 0.15 nm$^3$. However, this number is only a lower bound, since the full film is converted already for the lowest doses studied in the present work.

**CONCLUSIONS**

Exposure experiments with a low-energy electron microscope on Sn-based EUV resists, tin-oxo cages, allowed us to study the energy and doses of low-energy electrons that are required to render a solubility switch in this material. *In-situ* EELS spectroscopy proved essential to accurately determine the landing energy of the incident electrons on these insulating resist layers. *Ex-situ* AFM analysis of the exposed samples before and after a development process shows that electron exposure yields an insoluble material denser than the original resist. Moreover, prolonged exposure leads to further densification. This behavior is in agreement with sequential reactions induced by electron irradiation that we simplify as an **A→B→C** reaction model, where **B** represents an insoluble mixture of units cross-linked in low degree and **C** represents the subsequent formation of a more densified network resulting from a higher degree of carbon-loss and cross-linking of the SnOx inorganic fragments. Remarkably, electrons with energies as low as 1.2 eV can induce noticeable chemical changes in the resists. Furthermore, it was estimated that fewer than 10 electrons of 2-38 eV per molecule are necessary to render the solubility switch, which corresponds to an average reaction volume of 0.15 nm$^3$ per electron. In this energy range, higher-energy electrons do not cause significantly more conversion than low-energy electrons. The present work uses an unprecedented approach to evidence how crucial electrons of very low energy are for the solubility switch of EUV resists, even in low amounts. The insights gained in this investigation are of great value to the understanding of how inorganic EUV resists operate in lithographic applications.

**METHODS**

*The EUV Photoresist.* 6.5 x 6.5 mm$^2$ Boron-doped Si substrates (p-type) covered with a native oxide layer are used for preparation of photoresist thin films. The TinOH material is dissolved in toluene to a concentration of 7.5 mg/mL. Solutions were filtered through a 0.25-μm PTFE filter right before spin coating. TinOH thin films are obtained by spin coating under 2000 rpm for 45 s with a speed of 750 rpm/s on the piranha base-cleaned Si substrate. Details about materials synthesis and preparation can be found elsewhere.[28,29,47] The thickness of the resulting films is 20 nm as determined by AFM.

*Low Energy Electron Microscopy.* All LEEM/EELS experiments on measurement of electron landing energy and exposure of photoresist to low energy electrons described here are performed using the aberration-corrected ESCHER LEEM experimental setup (Leiden University) which is based on a commercial LEEM SPECS P90 instrument design. Details about the LEEM outline and the microscope capabilities can be found elsewhere.[33,34,36]

The microscope is operated at an electron gun energy of 15 keV. Before interaction with the sample, the 15 keV electrons emitted by the electron gun are slowed down to 0 - 40 eV energy ($E_0$) by negative biasing of the sample relative to the grounded objective lens. Specularly *reflected* and secondary electrons are extracted by this bias field and leave the sample with no possibility of return.

*Electron landing energy measurement.* Experiments on measurement of electron landing energies ($E_{land}$) are performed in LEEM using Electron Energy Loss Spectroscopy (EELS). The $E_{land}$ measurement principle is based on measurements of energy distribution of electrons emitted and/or reflected from the sample upon exposure of the surface to a primary electron beam of well-defined energy $E_0$ and current density ($I_0$). Upon interaction of the primary electron beam with the photoresist surface, depending on the $E_0$ value, specularly reflected and/or secondary electrons with energy $E_{land}$ leave the sample surface. After passing the electron optics system, beam separators and electron mirror, the reflected and/or secondary electron beam reaches the detector and the resulting image, representing an electron energy distribution spectrum in ($E$, $k_y$) space (see *Supporting Information*) is recorded using a micro-channelplate array and a CCD camera. All electron energy distribution spectra are corrected for detector-induced artifacts by subtracting a dark count image and their intensity is normalized before further analysis.

*Exposure to low-energy electrons in LEEM.* Exposure to electrons of well-controlled energy, current density and dose is performed using a beam blanking system. For each single exposure a value of $E_0$ is chosen in the range from 0 up to 40 eV. The value of $E_0$ is constant during each single exposure. When the exposure is finished, the beam is blanked and the sample is moved to a new unexposed position. After this the procedure is repeated in the same way, but with a different value of $E_0$. This approach creates a 2D array of exposed areas where one coordinate corresponds to change of dose at constant $E_0$, while the ortogonal axis corresponds to changes of $E_0$ at constant dose (see *Supporting Information*). After electron exposure in LEEM and AFM analysis of the as-exposed sample, this same sample is developed in 1:3 heptanone:water solution for 30 sec. The quality of the resulting patterns is checked using an optical microscope.

*Atomic Force Microscopy.* The analysis of the sample topography directly after exposure and after development is performed using a commercial Atomic Force Microscopy (AFM) instrument (Bruker). The microscope is operated in tapping mode. AFM micrographs are generated using commercial silicon and silicon nitride tips. Analysis of AFM micrographs is performed using Gwyddion software[48].

*Density Functional Theory calculations.* A model of TinOH was built starting from the crystal structure.[27] The isopropanol molecules were replaced by water molecules to reduce the computational cost. The geometry of the complex was optimized using the B3LYP hybrid functional with the LANL2DZ effective core potential basis set, using the Gaussian16 program.[49] The relatively small basis set was chosen in order to keep the size of the calculations manageable, but we also found that the structures obtained agreed better with experimental crystal structures than those optimized using the larger Def2SVP basis set. For better evaluation of the relative energies we used single point calculations with the Def2TZVP basis set (B3LYP/Def2TZVP//LANL2DZ). Bond dissociation energies were corrected for the differences in zero point vibrational energies (B3LYP/LANL2DZ).

**ACKNOWLEDGEMENTS**

Part of this work has been carried out within ARCNL, a public-private partnership between UvA, VU, NWO and ASML and was partly financed by "Toeslag voor Topconsortia voor Kennis en Innovatie (TKI)" from the Dutch Ministry of Economic Affairs. The authors are grateful to Marcel

Hesselberth and Douwe Scholma for their indispensable technical support. We also thank Joost Frenken (ARCNL) for insightful discussions and suggestions.